\documentclass[9pt]{emulateapj}         % Comment this line out
\usepackage{graphics} % for pdf, bitmapped graphics files
\usepackage{epsfig} % for postscript graphics files
\begin{document}

\title{Detection of fast radio transients with multiple stations: a case study using the Very Long Baseline Array}
\shorttitle{Detection of fast radio transients with multiple stations}
\shortauthors{Thompson et al.}
\email{david.r.thompson@jpl.nasa.gov}

\author{ 
David R. Thompson\altaffilmark{1}
Kiri L. Wagstaff\altaffilmark{1}, 
Walter Brisken\altaffilmark{2}, \\ 
Adam T. Deller\altaffilmark{2,3}, 
Walid A. Majid\altaffilmark{1}, 
Steven J. Tingay\altaffilmark{4}, 
Randall Wayth\altaffilmark{5}}

\altaffiltext{1}{Jet Propulsion Laboratory, California Institute of Technology, 4800 Oak Grove Dr., Pasadena, CA 91109.}
\altaffiltext{2}{National Radio Astronomy Observatory.  P.O. Box 0, 1003 Lopezville Road, Socorro, NM}
\altaffiltext{3}{University of California, Berkeley.}
\altaffiltext{4}{International Centre for Radio Astronomy Research / Curtin University. GPO Box U1987, Perth WA, 6845. Australia}

\begin{abstract}
Recent investigations reveal an important new class of transient radio phenomena that occur on sub-millisecond timescales.  Often transient surveys' data volumes are too large to archive exhaustively.  Instead, an on-line automatic system must excise impulsive interference and detect candidate events in real-time.   This work presents a case study using data from multiple geographically distributed stations to perform simultaneous interference excision and transient detection.   We present several algorithms that incorporate dedispersed data from multiple sites, and report experiments with a commensal real-time transient detection system on the Very Long Baseline Array (VLBA).  We test the system using observations of pulsar B0329+54.  The multiple-station algorithms enhanced sensitivity for detection of individual pulses.  These strategies could improve detection performance for a future generation of geographically distributed arrays such as the Australian Square Kilometre Array Pathfinder and the Square Kilometre Array.
\end{abstract}


\section{Introduction}

The radio sky varies over a wide range of timescales \citep{Cordes2004}.  Recent studies have characterized populations of {\it slow transient} radio sources that vary over timescales from seconds to years \citep{Bower2007,Croft2010,Bannister2010}.  Observers have also discovered an important new class of {\it fast transient} events at millisecond- to sub-millisecond timescales \citep{Lazio2009,Cordes2004}.  These include Gamma Ray Bursts \citep{Cameron2005}, Rotating Radio Transients (RRATs) \citep{McLaughlin2006}, and unique single-pulse phenomena like the Lorimer Burst \citep{Lorimer2007}.  Fast transients' short durations imply high-energy coherent processes, giving them significant scientific importance.  However, few surveys have specifically targeted fast transients, and with few validated detections these populations are poorly characterized.  The challenge has motivated considerable interest from the radio astronomy and pulsar communities, with several recent and forthcoming searches for transient signals in array time series data.  Such surveys include the ATA Fly's Eye \citep{VonKorff2009,Siemion2010}, the LWA transients study \citep{Taylor2006}, the LOFAR transient campaign \citep{Hessels2009}, and the ALFA pulsar search \citep{Deneva2009}.  In the near future, a new generation of instruments with significantly improved survey speed and sensitivity will begin operations.   The Square Kilometre array \citep{Cordes2003, Hall2008} and its precursor projects such as the Australian Square Kilometre Array Pathfinder (ASKAP) \citep{Johnston2008}, the Murchison Widefield Array (MWA) \citep{Lonsdale2009}, and MeerKAT \citep{Jonas2009} will open additional observational parameter space to new and unanticipated transient sources.

Any transient survey must demonstrate that detected events are not of terrestrial origin.  Frequency dispersion is strong evidence, but perhaps not conclusive proof since terrestrial events may mimic dispersion profiles \citep{Burke-Spolaor2010}.  Localizing a source within a calibrated image would provide more conclusive evidence, as would coherent dedispersion to resolve its temporal structure.  However these analyses require access to raw antenna voltages and infeasible data storage volumes.   Therefore investigators often buffer time series voltage data just long enough to quickly identify probable transients, and save the only the most promising data to archival storage for a full coherent analysis \citep{Macquart2010}.  Accurate candidate selection is essential because of limits to storage space and the time of human analysts to examine detections off-line \citep{Ellingson2004}.  Transient searches must address the {\it algorithmic} challenge of real-time event detection in incomplete, noisy data.
  
Transient searches in time series data are uniquely sensitive to impulsive disruptions from instrumental gain variations and Radio Frequency Interference (RFI).  Such phenomena have less effect on imaging studies since they generally disappear during correlation.  However, impulsive noise is similar in character to single pulse transients making it a significant practical challenge to real-time detection.  Investigators generally treat interference excision and source detection separately; they first remove contaminated segments and then detect candidate signals in the remainder.  Typical excision algorithms use indicators like atypical spectral kurtosis \citep{Deller2010}, the lack of frequency dispersion \citep{Deneva2009}, or the narrow bandwidth typical of artificial sources \citep{Bhat2005}.  Median filtering is often used to mitigate impulsive noise. In practice most arrays also use {\it ad hoc} rules for site-specific interference.  However, it is always difficult to excise interference entirely, and transient sources are so rare that occasional terrestrial signals easily dominate the effective sensitivity achievable for a given archiving budget.  Very Long Baseline configurations with distributed stations further increase exposure to hardware faults and RFI.  Future installations like the Square Kilometre Array will have dramatically larger scale and complexity but the number of human analysts for manual post analysis will remain relatively constant, making interference mitigation a vital enabling technology \citep{Ellingson2004}.  More generally, a principled approach to disambiguate interference will be important for validating any positive detections.

This work exploits geographic separation to adaptively and jointly classify interference, background noise, and novel transient signals.  In general, interference is statistically independent at widely-separated stations.   Detectors can exploit this principle to discriminate terrestrial events in real-time without computationally expensive  coherent analysis.  Geographic separation enables unambiguous classification of non-terrestrial sources, making very long baseline configurations especially valuable for fast transient surveys.  Most previous studies of the transient detection problem treat the single-dish or single-station case \citep{Fridman2010}. At least one other investigation has used dual station detection for RFI excision \citep{Bhat2005}.  Bhat et al. use two stations' independent detections, comparing the final event lists to create an RFI excision mask.  Our work explores the most general formulation of online joint RFI excision and source detection incorporating the detected signal strength at all stations simultaneously, for observations collected at many distributed locations.  

\begin{figure*}[]
\centerline{\epsfig{file=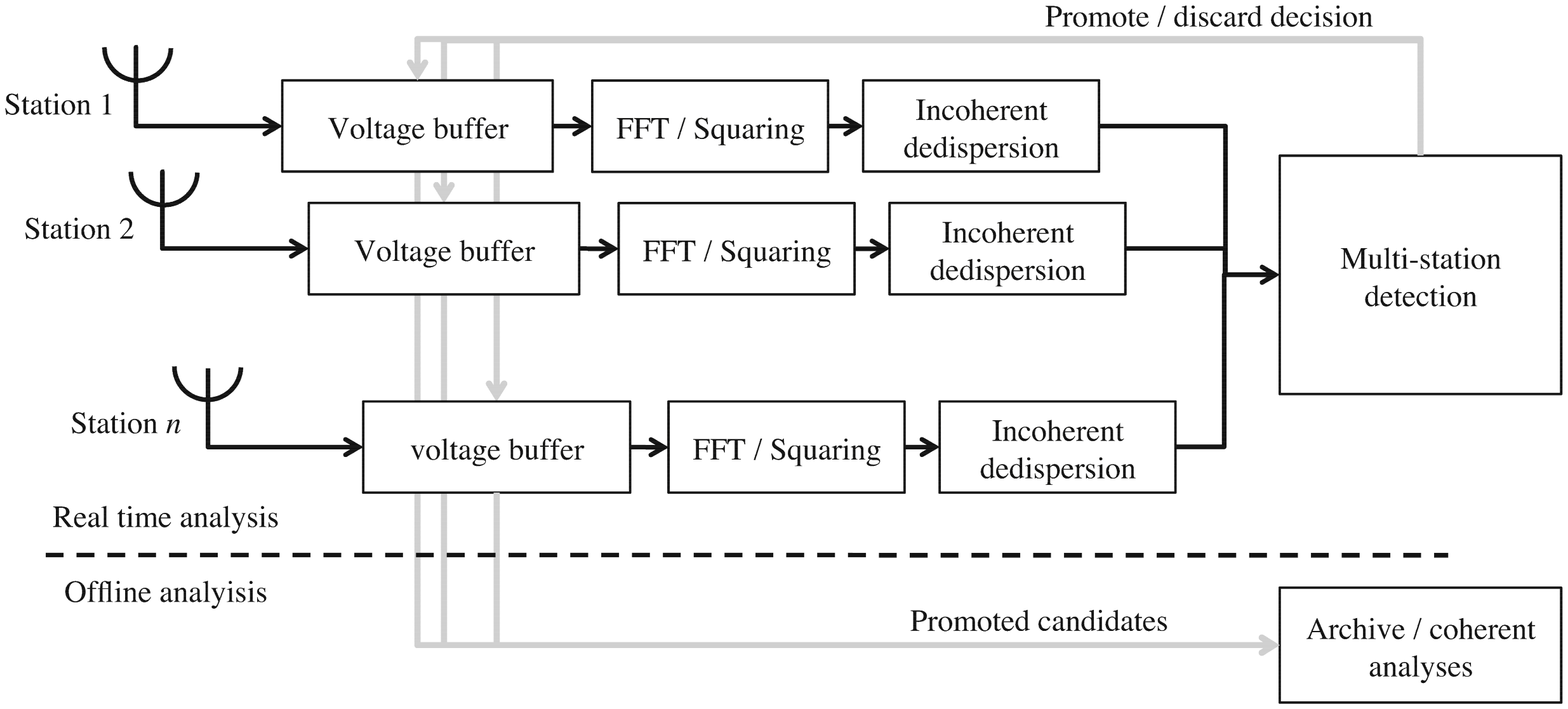, width=0.8\linewidth}}
\caption{Multi-station architecture for transient detection. An incoherent dedispersion search produces a separate time series for each station and candidate DM.  The detector stage analyzes the streams from all stations and makes real-time decisions about which time segments to promote.}  
\label{fig:architecture}
\end{figure*}

This work focuses on {\it incoherent} transient detection, where received signals are channelized in a spectrometer and squared, distinct from the more computationally expensive {\it coherent} approaches using phase information.  Figure \ref{fig:architecture} shows the basic components of a multiple-station incoherent transient detection system.  Here a set of $n$ stations observes some common source, and stores the complete raw voltage data to a rolling buffer.  This voltage data is then transformed by channelization and squaring into a matrix of incoherent power measurements at discrete frequencies and timesteps.   A transient detection system analyzes the $n$ independent data streams, searches for probable events, and triggers occasional transfers from the buffer into a permanent archive whenever it discovers a likely candidate.  Note that while the term ``detection'' is often used to describe the initial squaring operation, our use of the term always refers to the final promote/discard decision.

This paper describes several detection algorithms and characterizes their performance with respect to both detection sensitivity and resilience to RFI.  It then presents experimental trials using the software correlator of the Very Long Baseline Array (VLBA) \citep{Deller2007,Deller2010b}.  We describe V-FASTR, a commensal real-time detection system developed as a precursor to the Australian SKA Pathfinder Project \citep{Wayth2011}.  V-FASTR incorporates multiple stations and adapts to dynamic antenna configurations and RFI conditions.   Its observations of pulsar B0329+54 demonstrate sensitivity improvements from on line adaptivity and multiple-station synthesis.  This suggests that multiple-station algorithms might improve performance of commensal transient searches by other geographically distributed instruments, such as the future Square Kilometre Array.

\section{Detection methods}
\label{sec:methods}

\cite{Bhat2005b} classify transient detection strategies according to the number of independent beams and stations involved; {\it multiple-pixel} detection uses several fields of view while {\it multiple-station} detection uses several geographic locations that observe a common target.  Table \ref{tab:bhattaxonomy} expands their taxonomy with recent and anticipated transient detection projects.   This work deals with multiple-station transient detection, applicable to VLBI instruments as well as future installations such as ASKAP, MeerKAT, and the SKA.  The VLBA transient project is the first detection system excising RFI with more than two separate locations.  The following sections present several basic flavors of multiple-station algorithms, first establishing notation for the single-station case and then extending this framework to incorporate many geographic locations.

\begin{table}[]
\begin{centering}

\begin{tabular}{|p{2cm}|p{2.75cm}|p{2.75cm}|}
\hline
 & Single-Pixel & Multiple-Pixel \\
\hline
 Single-Station & 
   Arecibo WAPP, LWA \citep{Taylor2006}, LOFAR campaign by \cite{Hessels2009} & 
   ALFA \citep{Deneva2009}, ATA Fly's Eye \citep{VonKorff2009,Siemion2010} \\
\hline
  Dual-Station & 
   Arecibo and GBT \citep{Bhat2005} & 
   N/A\\
\hline
   Multiple-Station & 
   VLBA/V-FASTR, this work &
   ASKAP \citep{Macquart2010}, SKA \citep{Cordes2003} \\
\hline
\end{tabular}
\end{centering}
\caption{Taxonomy of incoherent transient detection methods based on \cite{Bhat2005}}
\label{tab:bhattaxonomy}
\end{table}

\subsection{Single-station detection}

We consider the output of a single station to be a function of both time and frequency, $S(t, \nu)$.  Here, $S$ could be either the voltage or, anticipating discussion below, the autocorrelation of the voltage.  More generally, the output could have other dependencies such as polarization.  We consider a function $f$ that operates on a subset of $S$, for instance, a short segment of the data in time or a restricted frequency range.  Formally we can represent each segment as a vector $\mathbf{x} \in \mathcal{X}$.  Here $\mathbf{x}$ is a multivariate data point comprised of the frequency and time values in a single segment.  We seek a single-valued {\it discriminant function} $f(\mathbf{x})$ from which transients can be detected in that $f(\mathbf{x})$ is large if there is an astronomical transient present and $f(\mathbf{x})$ is small if the data contain only noise or RFI.  Promising segments whose values exceed a user-defined threshold $\tau$ are promoted to archival storage and coherent analysis.

%A radio array produces discrete time series data where each timestep $t$ is associated with one or more radiometric observations of received power at frequencies $\nu$.  The transient detector classifies a fixed-size time interval whose data can be represented as the vector $\mathbf{x} \in \mathcal{X}$.  The detection rule is a discriminant function $f(\mathbf{x}): \mathcal{X} \mapsto \RR $ that should be large for time segments containing transient events and small for instrument noise or RFI.  A threshold $\tau$ determines which time intervals are promoted to archival storage and coherent analysis.
 
Transient signals are distorted by dispersion from their passage through the interstellar medium.  This manifests as a frequency-dependent time delay that is inversely proportional to the signal's frequency.  Following \cite{Lyne1998}:
\begin{equation}
\Delta t_{\rm delay}  = 4.1{\rm ms}\frac{DM~k}{\Delta\nu^2_{\rm GHz}}
\end{equation}
\noindent where $\Delta\nu$ is the observation bandwidth and $DM$ is the Dispersion Measure of the source, a quantity representing the integrated free electron density along the line of sight.  A broadband pulse experiences a different delay at each frequency and is thus distorted into a time-swept curve.
%{\bf [KLW: Include a figure such as Lorimer's pulse here?]}.   
Detecting broadband pulses requires a filter that matches this dispersion sweep in the time/frequency domain.  Equivalently, one can correct the delay independently in each frequency with {\em dedispersion}, and subsequently apply a simultaneous matched filter over all channels \citep{Bhattacharya1998,Deneva2009}. Typically the DM is not known in advance, but searching over a set $\mathcal{D}$ of DM values provides sensitivity to many possible dispersion profiles.  Typical observed DMs range from $0$ for terrestrial events, up to order $10^2$ for local sources, to order $10^3$ for sources near the galactic center where the interstellar medium is dense.  Negative dispersion measures do not correspond to any anticipated natural phenomenon, but they may also be tested for false detection statistics and relevance for Extra Terrestrial Intelligence (ETI) investigations.  The optimal DM search spacing is related to the frequency range, filterbank channel width and time resolution \citep{Cordes2003}.   DMs can be searched in parallel so the transformation is amenable to multi-core software solutions. Other methods for real-time dedispersion include GPU or FPGA processing \citep{VonKorff2009} or efficient caching structures such as Taylor trees \citep{Taylor1974}. 

We use the operator $\phi(\mathbf{x},d)$ to signify a matched filter shaped for dispersion to a specific DM $d$.  The detection decision can be independent for each dedispersed segment, leading naturally to the classical maximum likelihood discriminant function:
\begin{eqnarray}
{\rm promote~ if ~}  f(\mathbf{x}) > \tau \nonumber\\
{\rm \quad for} \quad f(\mathbf{x}) = \displaystyle\max_{d \in \mathcal{D}} \phi(\mathbf{x},d) \label{eqn:onestation}
\end{eqnarray}
\noindent Disregarding interference, in the ideal case both sky and instrument noise in the time domain are gaussian-distributed.  After squaring and integration the summed samples follow a Chi-squared distribution $\chi^2_m$ with many degrees of freedom, which we can approximate by another gaussian.  This leads to the following expression for the minimum detectable {\it intrinsic} peak flux density \citep{Bhattacharya1998}:
\begin{eqnarray}
S_{min} &=& \left(\frac{\tau}{\sigma}\right)\frac{S_{sys}}{\sqrt{N_{pol} \Delta \nu \Delta t}} \label{eqn:sens}
\end{eqnarray}
\noindent Here $S_{sys}$ is the system-equivalent flux density, $N_{pol}$ is the number of polarizations, $\Delta \nu$ is the total bandwidth of the filter, and $\Delta t$ is the intrinsic pulse duration.  The term $\left(\frac{\tau}{\sigma}\right)$ is simply the SNR detection cutoff (typically 5), with $\sigma$ the standard deviation of $f(\mathbf{x})$ for noise.  One can easily calculate sensitivities for a known false alarm rate and receiver sensitivity.  

This sensitivity estimate is invalid when additive interference makes the detector output non-gaussian.  We model occasional impulsive RFI by a distribution $\mathcal{G}_0$, and a real transient source by $\mathcal{G}_1$. For the hypothesis $H_0$ that the segment is noise, and $H_1$ that the event is a transient, we have: 
\begin{eqnarray}
\mathbf{x} | H_0 \sim \chi^2_m + \mathcal{G}_0 \quad {\rm and} \quad \mathbf{x} | H_1 \sim \chi^2_m + \mathcal{G}_1 \label{eqn:model}
\end{eqnarray}
\noindent Most segments do not contain RFI so $\mathcal{G}_0$ has a large probability mass at $0$.  However, even occasional interference can quickly dominate detections making it a major practical impediment to the effective survey yield.  Therefore, in addition to quantifying a detector's flux sensitivity it is also useful to examine its power as a {\it classifier}, i.e. its ability to discriminate between true events and interference.  

We measure classification performance with a quantity from decision theory known as the expected loss $E[\mathcal{L}]$.  This incorporates the cost $\mathcal{L}_{FP}$ of any false positive detections and the cost $\mathcal{L}_{FN}$ of all false negatives:
\begin{eqnarray}
E[\mathcal{L}] &=& \int_\mathbf{x} \mathcal{L}_{FP}~p(f(\mathbf{x}) \geq \tau)~p(\mathbf{x}|H_0) p(H_0) d\mathbf{x} + \nonumber \\
               &&\int_\mathbf{x} \mathcal{L}_{FN}~p(f(\mathbf{x}) < \tau)~p(\mathbf{x}|H_1)p(H_1) d\mathbf{x} 
\end{eqnarray}
\noindent We collapse unknown but static terms into constants $\alpha_1$:
\begin{eqnarray}
E[\mathcal{L}] &\propto& \alpha_0~p(f(\mathbf{x})\geq \tau | H_0) + \alpha_1~p(f(\mathbf{x}) < \tau | H_1)\nonumber\\
               &\propto& \alpha~p(f(\mathbf{x})\geq \tau | H_0) + p(f(\mathbf{x}) < \tau | H_1) \nonumber
\end{eqnarray} 
\noindent This reduces under a monotonic transform to a weighted sum of the {\it false positive rate} and {\it true positive rate}. 
\begin{eqnarray}
E[\mathcal{L}] &\cong& \alpha~p(f(\mathbf{x})\geq \tau | H_0) - p(f(\mathbf{x}) > \tau | H_1) \label{eqn:loss}
\end{eqnarray} 
\noindent A joint detection and excision rule is simply a discriminant function $f(\mathbf{x})$ that aims to optimize this objective.  
 
\begin{figure}[]
\centerline{\epsfig{file=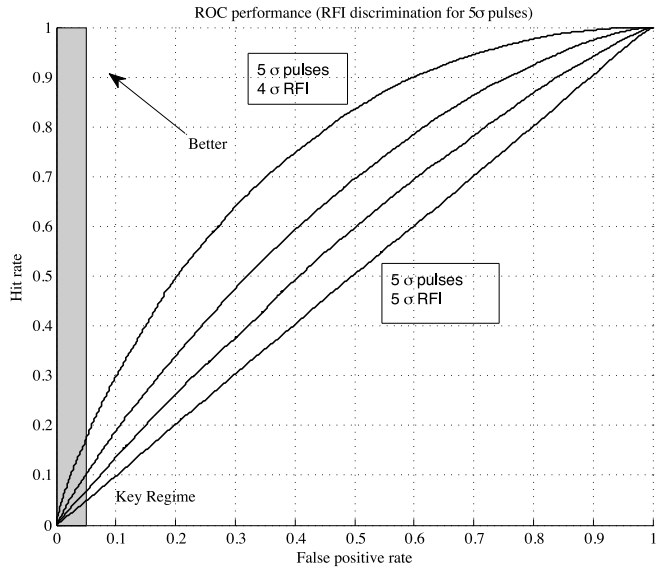, width=\linewidth}}
\caption{ROC curve showing classification performance on a simulated dataset, for single-station detection under different RFI environments.  }  
\label{fig:singlestation}
\end{figure}

One can compare classification performance under different cost assumptions using a Receiver Operating Characteristic (ROC) curve such as Figure \ref{fig:singlestation}. The ROC plot shows all possible true and false positive rates for different choices of $\tau$.  For a given possible false alarm tolerance (i.e., a particular $\alpha$ value in equation \ref{eqn:loss}), the best performance achievable for the discriminant lies somewhere on the ROC line.   A random detection rule, which provides no information about the RFI/transient distinction, corresponds to a diagonal line with equal false positive and true positive rates.  Better detection results approach the upper left region.   

This example shows a simulation where a hypothetical time series is dedispersed to the correct $DM$ for each of $32$ independent frequency channels so that no residual time delay remains.  Matched filtering yields a $\chi^2$ background noise signal.   We synthesize a dataset of 10000 timesteps with additive RFI and additive transient pulses in equal proportion.  Both kinds of events have a single timestep width, and a constant SNR is used so that the only random element is measurement noise.   If RFI pulses are weaker on average than transients, then the detection rule favors true transient events and the curve approaches the upper left.

The area under the ROC curve is commonly used as a figure of merit to summarize the ROC performance.  However, for the transient detection task, very few candidates can be promoted so the majority of this area is not relevant.  Instead, the most important aspect of performance is the ROC in the regime of extremely low false positive rates.  For our observational study that follows we will consider the area under the curve for a false positive rate of $<0.01$, corresponding to a detection rule that promotes $1\%$ of all time segments.  ROC analysis, together with sensitivity under perfect observing conditions (e.g. equation \ref{eqn:sens}), permits principled comparisons of single- and multiple-station detection strategies. 

%Note that when RFI and transient pulses are equal in strength the single-station detection rule performs no better than a random selection.  If RFI pulses are weaker on average than transients, then the summed intensity helps to classify true transient events.   Note that it is only feasible to promote a very small fraction of timesteps for in-depth analysis, so we are most concerned with the leftmost portion of the graph corresponding to very low false positive rates. 

\subsection{Multiple-station detection algorithms}

In the general multiple-station case with $n$ geographic locations signals are dedispersed and filtered independently for each station and DM.  This gives a combined data set of size $|\mathcal{D}| \times n$ at every timestep.  Geographic separation can assist with detection because transients are correlated across multiple stations, while (local) RFI is not.  Here we present several alternative families of multi-station detection algorithms.  They vary in computational complexity, accuracy, and ease of implementation.  For simplicity we assume that all receivers have similar sensitivity, though multiple-station methods could also benefit more diverse systems such as LOFAR and the SKA that have centralized concentrations of collecting area.

\subsubsection{Sum of Signals}

\begin{figure}[]
\centerline{\epsfig{file=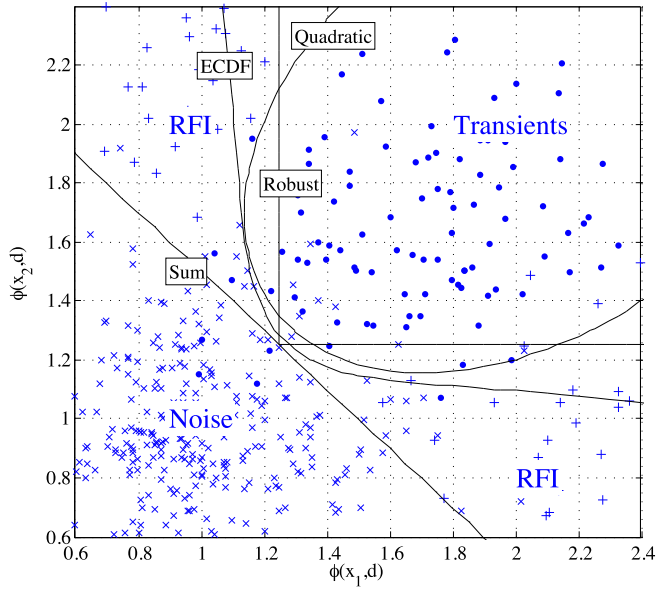, width=\linewidth}}
\caption{``Decision boundaries'' of multiple-station detection methods, illustrated in the simplest case of two stations.  The axes correspond to the signal strength at each station.  RFI yields a strong response at one location, while actual transient signals appear at multiple stations simultaneously.}  
\label{fig:decisionboundaries}
\end{figure}

For incoherent detection with a matched filter, one achieves maximum sensitivity by summing the dedispersed signal over all stations as in \cite{Hessels2009}.  The discriminant function is the maximum of all such sums at each DM:
\begin{eqnarray}
f_{sum}(\mathbf{x})&=&\displaystyle\max_{d \in \mathcal{D}} \frac{1}{n} \displaystyle\sum_{a=1}^n \phi(\mathbf{x}_a,d)
\end{eqnarray}
\noindent For $n$ stations this rule provides $\sqrt{n}$ improvements in detection sensitivity.
\begin{eqnarray}
S_{sum} &=& \left(\frac{\tau}{\sigma}\right)\frac{S_{sys}}{\sqrt{N_{pol} \Delta \nu \Delta t~n}} \label{eqn:sens_sum}
\end{eqnarray}
\noindent   Figure \ref{fig:decisionboundaries} portrays this detection strategy.  It shows a simplified case with RFI and transient signals for just two stations.  The axes show stations' matched filter responses in arbitrary units after dedispersion to the appropriate DM. Scattered points show noise, RFI-contaminated, and true transient segments drawn from the basic model.  We use unrealistic proportions of event types in order to show a significant number of each.  RFI appears in just one station at a time, so these points lie close to the axes.  Transient events show large signals at all stations.    

The summation rule corresponds to a discriminant function with hyperplane isocontours.  Any specific choice of detection threshold $\tau$ defines one isocontour as the {\it decision boundary} separating detected and rejected data.   Figure \ref{fig:decisionboundaries} shows a typical decision boundary  with the line labeled {\it Sum}.   This illustration demonstrates why summation can never capture all transient events without also including some interference.  Geographically separated stations could actually magnify this effect, since each additional geographic location brings a new RFI environment.  

\begin{figure}[]
\centerline{\epsfig{file=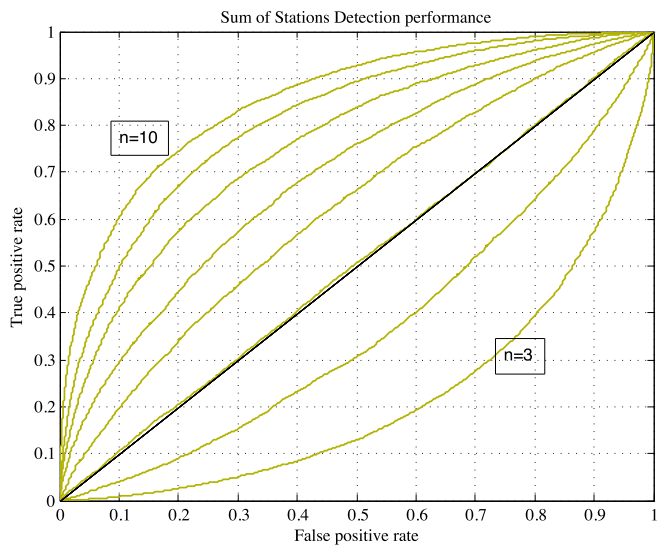, width=\linewidth}}
\caption{ROC curve showing false positive and true positive rates for discriminating weak transients from impulsive RFI.  The simulation uses the basic summation rule as a discriminant.  The data set consists of 30,000 timesteps containing an equal number of transients (SNR 1--2) and RFI events (SNR 5-10).}  
\label{fig:roc_sum}
\end{figure}

A simple simulation demonstrates the summation rule's performance with different numbers of stations.  
%This simplified example provides a controlled test to characterize the different conditions contributing to algorithms' performance, with an eye toward the VLBA application described later in Section \ref{sec:vlba}.  Our basic source and RFI model is identical to the single-station case.
Transients appear with equal magnitude at all stations, such that each stations' signal taken individually has SNR ranging randomly and uniformly from 1 to 2.   RFI events have SNR ranging from 5 to 10, with all power concentrated at one random station.   This scenario discounts several kinds of interference that are more difficult to model, such as periodic or switched-mode interference patterns.    We simulate 30000 timesteps with equal portions of transient and regular events.  The ROC curves in Figure \ref{fig:roc_sum} show detection performance as an RFI/transient classifier.  Not surprisingly, the classification power depends purely on the total signal so it must accumulate several stations before it outperforms random selection.   A coherent detection system summing voltage values instead of accumulated powers could have different performance characteristics.

\subsubsection{RFI Masks}

\begin{figure}[]
\centerline{\epsfig{file=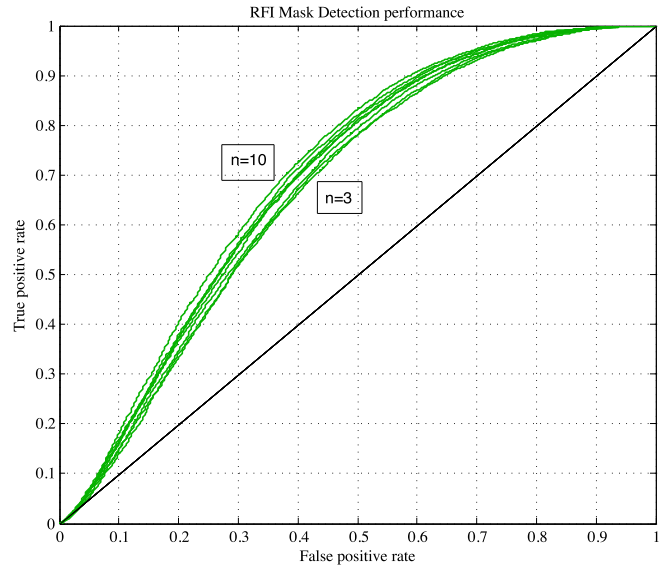, width=\linewidth}}
\caption{ROC curve for the RFI mask method.}  
\label{fig:roc_mask}
\end{figure}

\cite{Bhat2005} demonstrate a dual-station algorithm for RFI excision during joint observations at Arecibo and the Green Bank Telescope.  They threshold the signals at each station independently and compare the resulting event lists to yield an RFI mask.  We generalize this approach to the many-receiver case by masking all events not detected in at least two stations:  
\begin{eqnarray}
f_{mask}(\mathbf{x})= \displaystyle\max_{d \in \mathcal{D}} \left\{
\begin{array}{crl}
0 & {\rm if} & q(\mathbf{x},d)\leq 1 \\
\displaystyle\max_{a} \phi(\mathbf{x}_a,d) & {\rm if} & q(\mathbf{x},d)>1
\end{array} \right.
\end{eqnarray}
The number of stations' signals exceeding $\tau$ is given by:
\begin{eqnarray}
q(\mathbf{x},d) = |\mathbf{x}_a : ~~\phi(\mathbf{x}_a,d)>\tau_a, ~~1 \leq a \leq n |  
\end{eqnarray}
The stations might have dissimilar RFI environments, prescribing a different threshold $\tau_a$ for each.  This masking operation provides near-perfect RFI excision. However, any detection of a real transient must occur {\it independently} in each receiver.  Therefore the minimum flux sensitivity is identical to the single-station case.  We have:
\begin{eqnarray}
S_{mask} &=& \left(\frac{\tau}{\sigma}\right)\frac{S_{sys}}{\sqrt{N_{pol} \Delta \nu \Delta t}} \label{eqn:sens_mask}
\end{eqnarray}
Figure \ref{fig:roc_mask} shows ROC performance as an RFI/transient classifier for varying numbers of stations.  Additional stations do not improve performance; the overall detection sensitivity of the system is always equal to the second most sensitive station.

\subsubsection{Robust Sum}

Signals from geographically distributed stations are tantamount to Independent and Identically Distributed ({\sc IID}) samples from a common univariate process.  From this perspective RFI events are outliers that can be mitigated with robust estimation strategies.  Examples include trimmed and Winsorized estimators \citep{Leonowicz2005}.  The two-tailed trimmed estimator removes the strongest and weakest $k$ stations to produce the following discriminant function.  With stations ordered by signal strength,
\begin{eqnarray}
f_{robust}(\mathbf{x})&=&\displaystyle\max_{d \in \mathcal{D}}  \frac{1}{n-2k}\displaystyle\sum_{a=k+1}^{n-k-1} \phi(\mathbf{x}_a,d)
\end{eqnarray}
\noindent  The two-tailed trimmed estimator requires at least three stations, but the one-tailed version (which simply excises the strongest signal)  gives a meaningful result for just two stations.   They produce axis-parallel decision boundaries like the line labeled {\it Robust} in Figure \ref{fig:decisionboundaries}.

Trimming stations incurs a sensitivity penalty.  One can characterize sensitivity using the empirical standard deviation $\hat\sigma_{robust}$; the corresponding gaussian gives the absolute signal strength at a given rejection threshold.    
\begin{eqnarray}
S_{robust} &=& \left(\frac{\tau}{\hat\sigma_{robust}}\right)\frac{S_{sys}}{\sqrt{N_{pol} \Delta \nu \Delta t}} \label{eqn:sens_robust}
\end{eqnarray} 

\begin{figure}[]
\centerline{\epsfig{file=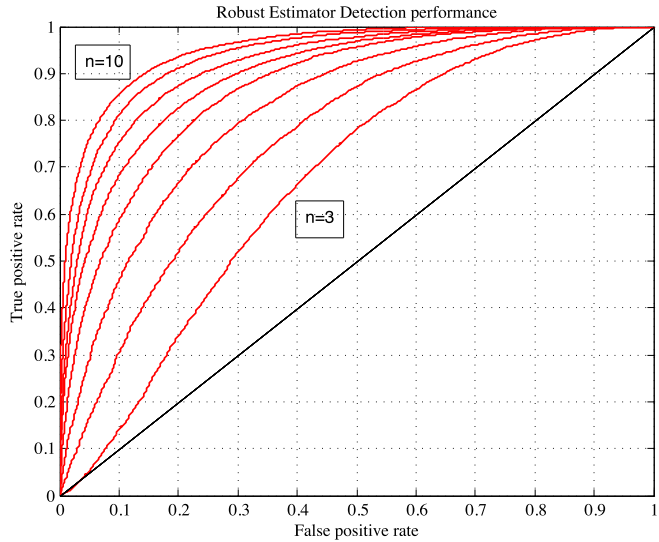, width=\linewidth}}
\caption{ROC curve for the robust estimator.}  
\label{fig:roc_robust}
\end{figure}

Figure \ref{fig:roc_robust} shows the robust estimator's ROC performance for varying numbers of stations.  In the three-station case performance is equal to the RFI masking approach since both methods use the second-strongest signal.  For four or more stations the robust estimator sums multiple measurements, increasing flux sensitivity and improving performance.

The trimmed decision rule requires signals to be sorted for each timestep and dispersion measure.  This operation is computationally tractable for existing very long baseline configurations.  A separate issue is the choice of the number of stations $k$ to trim at each timestep.  The best $k$ minimizes the expected loss from Equation \ref{eqn:loss}, which depends on data quality and the likelihood of simultaneous RFI events in multiple stations.  Ideally most timesteps exhibit ``clean'' gaussian noise, with RFI appearing only occasionally and in one station at a time.  In this case $k=1$ removes the RFI while maintaining maximum sensitivity to weak pulses.  If interference is not perfectly independent or if a receiver suffers from persistent noise conditions then simultaneous interference events could occur.  These situations would benefit from setting $k>1$.  If noise conditions change, one can find the current best setting for $k$ on-line by injecting synthetic pulses into the data stream and then attempting to detect them using several trimming values.

\subsubsection{Ensemble Estimate of CDF}

The {\it Ensemble CDF} (ECDF) rule mitigates RFI with a monotonic transformation of signal strengths that reduces the influence of extreme values from any single station.  Specifically, it estimates the probability that an observed signal exceeds a random typical timestep whose magnitude is the random variable $X$:
\begin{eqnarray}
f_{ecdf}(\mathbf{x})&=& \displaystyle\max_{d \in \mathcal{D}}\hat{F}(\phi(\mathbf{x},d)) \nonumber \\
        &=& \displaystyle\max_{d \in \mathcal{D}}\hat{p}(\phi(X,d) < \phi(\mathbf{x},d)) \nonumber \\
        &=& \displaystyle\max_{d \in \mathcal{D}} \frac{1}{n}\displaystyle\sum_{a=1}^n p(\phi(X_a,d) < \phi(\mathbf{x}_a,d)) \label{eqn:ecdf}
\end{eqnarray}
\noindent Each station maintains a separate probability estimate $\hat{p}$, so strong signals have less influence at stations with chronic RFI.   The expectation of this probability over all stations constitutes an ensemble estimate of the Conditional Density Function (CDF).  The method is reminiscent of the “mean rule” for combining multiple one-class classifiers, first suggested by \cite{Tax2001}.  It differs in that we are concerned only with high-intensity signals so we use the CDF in place of the standard density function.

One can compute the probability estimates $\hat{p}$ using any statistical density estimator, in advance from historical data or online from the current time series.  One practical on-line approach is to maintain an ordered list of recent signal strengths.   A binary search finds the percentile rank of a new observation, which provides an empirical CDF value \citep{Wasserman2006}.  Updating the ordered list requires an $O(\log n)$ insertion operation.   If constant-time computation is desired, one can discretize the CDF into $k$ unique values and store just $k$-tile signal strengths.  One can also reduce the computational burden by using a single probability estimate for multiple DMs. 

It can be shown that the ensemble estimator retains optimal flux sensitivity for detecting weak sources.  The discriminant rule is a sample average of a CDF which is concave wherever values are larger than the average noise.  Jensen's rule can be used to show that sensitivity is preserved in this region of interest.  A demonstration appears in the Appendix.  In brief, the CDF of the on-source mean remains constant in expectation, while the CDF of off-source RMS preserves $\sqrt{n}$ improvements in noise variance.  Absent practical concerns about discretization or accuracy in extreme tail regions, the rule maintains sensitivity to the weakest signals.
\begin{eqnarray}
S_{ecdf} &=& \left(\frac{\tau}{\sigma}\right)\frac{S_{sys}}{\sqrt{N_{pol} \Delta \nu \Delta t n}} \label{eqn:sens_ecdf}
\end{eqnarray}

%{\bf [KLW: In the list case presumably one must also remove the oldest signal strength unless the list is to grow in an unbounded way.  I can think of a constant-time way to achieve this, so it doesn't impact the log-n insertion time, but maybe the details of how one would do this should be included here?  Or not, if you advocate the discretization route?]}

\begin{figure}[]
\centerline{\epsfig{file=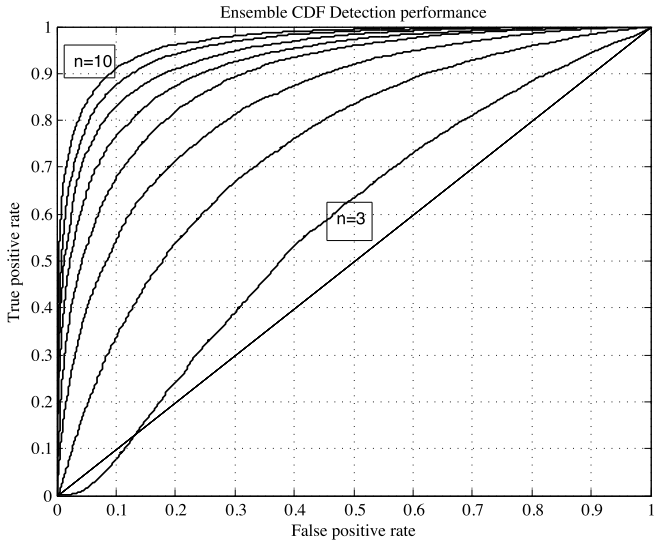, width=\linewidth}}
\caption{ROC curve for ensemble CDF discriminant.}  
\label{fig:roc_ecdf}
\end{figure}

Figure \ref{fig:roc_ecdf} shows ROC performance for the ensemble CDF function with varying numbers of stations.  It underperforms with just a few stations, but soon overtakes the robust estimator as the number of stations increases past four.  A potential advantage of the ensemble approach is that computing an independent probability score for each station compensates automatically for any systemic differences in their background signals or noise environments.  Finally, Figure \ref{fig:decisionboundaries} shows a typical decision boundary.  Transforming all signals to the $[0,1]$ interval down-weights the most extreme signal values and improves RFI rejection. 

\subsubsection{Quadratic Discriminant}

With examples of both transient and non-transient phenomena the detection problem becomes a traditional {\it supervised classification} task to find an optimal decision boundary separating two labeled data classes.  Typical machine learning solutions include Neural Networks or Support Vector Machines \citep{Bishop2006}.  Here we explore a simple quadratic decision boundary, which is optimal if on-source and off-source distributions can be characterized by multivariate gaussian PDFs.  The discriminant function is defined by a mean $\mu$ and a positive definite covariance matrix $\Sigma^{-1}$.   
\begin{eqnarray}
f_{quad}(\mathbf{x}) = \displaystyle\max_{d \in \mathcal{D}} (\phi(X,d)-\mu)^T~ \Sigma^{-1} ~(\phi(X,d) - \mu) 
\end{eqnarray}
\noindent  This supervised method can potentially achieve superior performance due to strong assumptions about the statistical properties of the source.  It accounts explicitly for sources' strengths and optimizes its decision boundary to the precise level of noise and degree of correlation across stations.   

\begin{figure}[]
\centerline{\epsfig{file=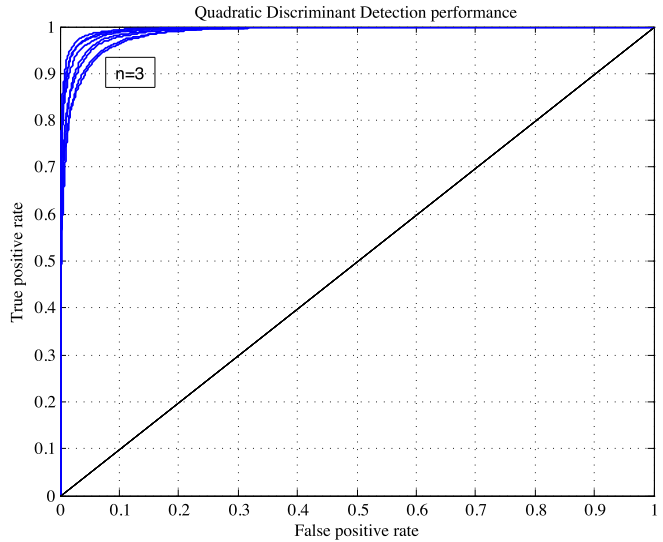, width=\linewidth}}
\caption{ROC curve for quadratic discriminant.  Supervised methods can provide superior performance if the training distribution is similar to the true data.}  
\label{fig:roc_quadratic}
\end{figure}

If its training assumptions are satisfied the sensitivity of the quadratic discriminant function is at least as good as the standard summation.  For example, in the RFI-free case, both noise and pulse distributions are truly gaussian with equivalent diagonal covariance matrices --- only the means differ.  Here the optimal decision boundary separating the two classes is a hyperplane.  More generally, the quadratic discriminant is optimally-sensitive as long as the data satisfies its assumption of gaussian-distributed classes.  In these cases a non-diagonal quadratic form is the proper likelihood ratio.  Figure \ref{fig:roc_quadratic} shows ROC performance.  

\begin{eqnarray}
 S_{quad} &=& \left(\frac{\tau}{\sigma}\right)\frac{S_{sys}}{\sqrt{n N_{pol} \Delta \nu \Delta t}} \label{eqn:sens_quad}
\end{eqnarray} 

Table \ref{tab:total_performance} summarizes the Area Under the ROC Curve (AUC) score for each of the discriminant functions and numbers of stations.  By this measure, the quadratic discriminant consistently outperforms all alternatives; its ability to discriminate impulsive RFI using just three stations is superior to the best performance offered by the next-best rule, for ten stations.  Its discrimination performance would be worse for new RFI environments or transients that did not match the training distributions.   In general it is best to estimate any detection method's free parameters on-line using the most current data.   

\begin{table}[]
\begin{centering}
\begin{tabular}{|l|c|c|c|c|c|c|c|c|}
\hline
\bf Method & n=3 &  4 & 5 & 6 & 7 & 8 & 9 & 10\\
\hline
\hline
\rm Sum &  .224  &  .367  &  .500  &  .610  &  .702  &  .756  &  .815  &  .856 \\
\hline
Mask       &  .676  &  .669  &  .679  &  .688  &  .692  &  .691  &  .700  &  .704 \\
\hline
Robust  & .676  &  .755  &  .818  &  .864  &  .899  &  .917  &  .937  &  .951 \\
\hline
ECDF  & .580 & .746  &  .833  &  .885  &  .923  &  .938  &  .955  &  .967 \\
\hline
Quad      &  .985  &  .985  &  .988  &  .989  &  .991  &  .993  &  .994  &  .995 \\
\hline
\end{tabular}
\end{centering}
\caption{Area Under the ROC Curve (AUC) score for each method, for various numbers of stations in simulated trials.}
\label{tab:total_performance}
\end{table}

\begin{figure}[]
\centerline{\epsfig{file=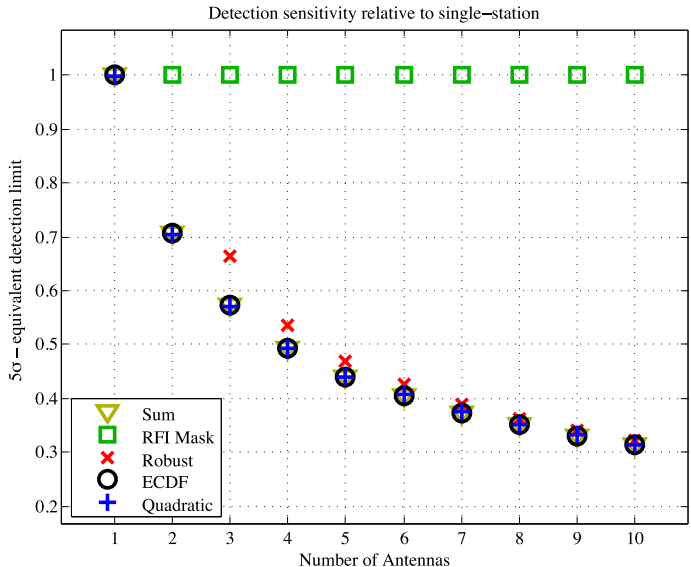, width=1.05\linewidth}}
\caption{Sensitivity for each method relative to single-station detection, in terms of noise RMS (lower is better) as the number of stations increases.}  
\label{fig:sensitivity}
\end{figure}

 Figure \ref{fig:sensitivity} shows sensitivity to weak pulses for each method in units of flux intensity relative to a single station. The robust method's sensitivity is difficult to describe analytically so we estimate it from the on-source mean and off-source RMS using 10000 timesteps of simulated data.  Its weak signal sensitivity is nearly indistinguishable from the classical summation rule for configurations with five or more stations.  This suggests that some form of robust estimation is almost always beneficial.  A conservative decision to excise just one or two stations from the sum causes the {\it smallest} marginal sensitivity impact but produces the {\it largest} marginal improvements in interference excision.  

Monte Carlo simulations can determine the sensitivity of more complex classifiers; one can fit a mapping from the detection probability onto signal strength using a function approximator like a smoothing spline.  We use this approach for the quadratic discriminant in Figure \ref{fig:sensitivity} and verify that sensitivity is indeed equivalent to the optimal summation rule.  Most methods in the diagram have similar sensitivity because it describes the RFI-free incoherent detection limit for perfect Gaussian noise.  Here a $ \sqrt{n}$ improvement is the best one can achieve.

\section{VLBA Observations}
\label{sec:vlba}

This section describes a case study of multiple-station transient detection using the Very Long Baseline Array (VLBA).  The experiment is part of the V-FASTR project, a trailblazer for the Australian Square Kilometre Array Pathfinder's CRAFT fast transients investigation \citep{Macquart2010}.  V-FASTR has installed a transient detection pipeline for commensal operation alongside standard VLBA observations.  A complete description of the architecture and initial results are provided in a companion paper \citep{Wayth2011}.  For completeness we will also provide a brief overview here.  

The VLBA has 10 geographically dispersed stations, each providing a single $25$m antenna.  These antennas are distributed across North America. The longest baseline stretches from Mauna Kea, Hawaii to Hancock, New Hampshire but the highest concentration of stations is in the Southwestern United States.  No two stations are with each other's local horizon, and anywhere from 2 to 10 stations may participate in an observation.  Voltage recordings are saved to disks and shipped to a central facility in Socorro, NM, where a computing cluster running the DiFX software correlator \citep{Deller2007,Deller2010b} processes signals for imaging and post-analysis.  

DiFX has been reconfigured to calculate auto-power spectra for each antenna, producing integrated frequency-channelized power measurements every $1$ ms.  An incoherent software dedispersion algorithm processes each station independently as in the architecture diagram of Figure \ref{fig:architecture}.  This stage uses three commercial multicore processors in parallel, and easily processes hundreds of dispersion measures in real-time.  Our tests to date have used 190 distinct dispersion measures while consuming just 10-20$\%$ of the total system capacity.  After dedispersion a transient detection stage processes the resulting time series and saves a small portion of the raw voltage data.   Online processing is essential since any archiving decisions must be made before the correlation job finishes and the entire disk is erased for reuse.

\subsection{Method}

\begin{figure}[]
\centerline{\epsfig{file=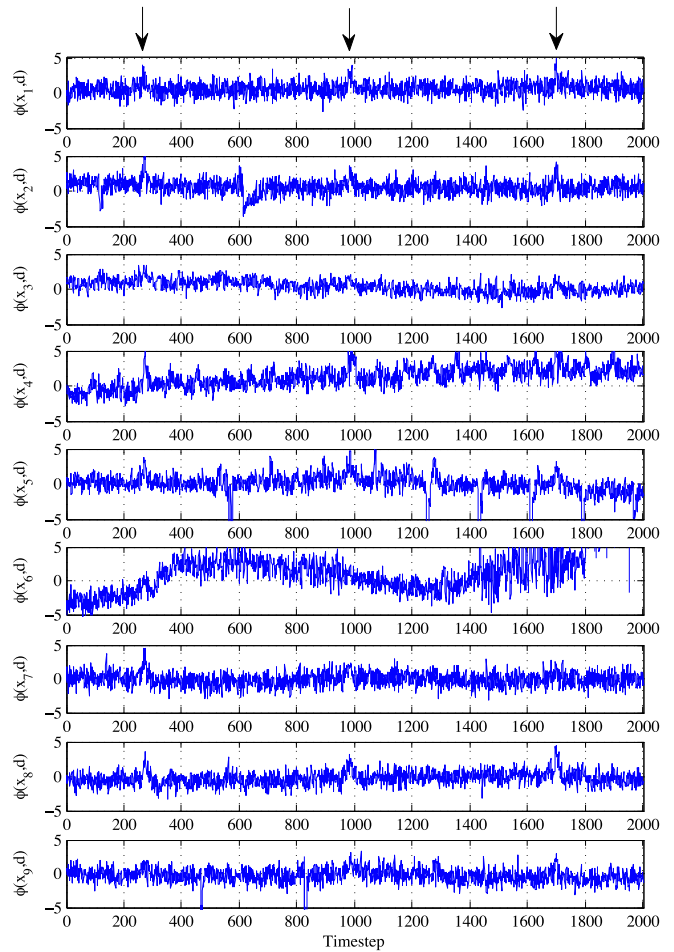, width=\linewidth}}
\caption{A typical segment taken from the third scan of the B0329+54 observation.  The actual pulses from the pulsar are indicated by arrows at the top of the diagram.}  
\label{fig:nineantennas}
\end{figure}

The pulsar B0329+54 was observed at four $8$MHz-wide bands evenly spaced from $1.4$GHz to $1.674$GHz; each band was channelized into $0.25$MHz frequency resolution and the resultant power spectra accumulated for  $1$ms.  This frequency resolution is typical for VLBA observations, and we use it here for fidelity to a commensal system.  After dedispersion to its known DM of $~26.8$pc$/$cm$^3$, the pulsar has an intrinsic pulse width of approximately $10$ms and is easy to resolve at this time resolution.  The pulsar period is approximately $714$ms; after dedispersion, typical data appears like the segment shown in Figure \ref{fig:nineantennas}.  This segment shows diverse interference including impulsive noise and systemic changes in the background at individual antennas.  Such interference would probably not significantly impact the cross-correlated measurements for which the VLBA system was originally designed, but it is problematic for finding short-duration events in the high-resolution time series data.

The pulsar was observed simultaneously at $9$ stations over 6 contiguous observation segments, or {\em scans}, with durations of $242$s.  Each scan was interspersed with measurements of a calibrator source; scans were spaced at approximately 5 minute intervals.  The observed pulse strength changed during the observation sequence, with signals weakening progressively in later scans.  Several antennas' average signal strengths also drifted slightly.  We compensated with a conservative high pass filter, subtracting the 100 timestep moving average from each observation. 

We assembled an authoritative event list by exploiting pulsar periodicity.  We posit that the actual events' arrangement has just two free parameters: start time and period.  We initialize these parameters by identifying the strongest pulses with visual inspection, and then extrapolate the other pulses' locations from the periodicity given by the catalog.  A precise optimization of start time and period then centered the events in the pulses by maximizing the mean signal over all events.  The typical SNR was $16$ after summing all antennas.  We created a test data set using positive examples drawn from the center of each pulse, and negative examples drawn at intervals between pulses.  We spaced the negative examples regularly at $10\%$ of the pulsar period, which provided a large sample but left enough separation between pulses to preserve statistical independence and insulate the negative samples from finite-width pulses.  We labeled timesteps as positive if they occurred at the right time according to the pulsar's rotational ephemeris, regardless of the actual received SNR.  This created a more difficult classification challenge and characterized discriminant algorithms' performance across many pulse strengths.  Each scan contained approximately $380$ pulse events and $3800$ negative examples.

We reserved the initial scan for training.  This training scan was effectively scan 0, and we will omit it from performance reports.  We used the remaining scans (1 through 5) to evaluate the detection algorithms.  We also computed algorithms' performance for all five test scans combined.  We set all free parameters through optimization on the data from the training scan, using the value $k=2$ for the robust estimator.  The Ensemble CDF algorithm did not require prior training; instead we estimated the CDF using the data from each scan in progress using a nonparametric plug in method \citep{Wasserman2006}.

\subsection{Results}

\begin{figure}[]
\centerline{\epsfig{file=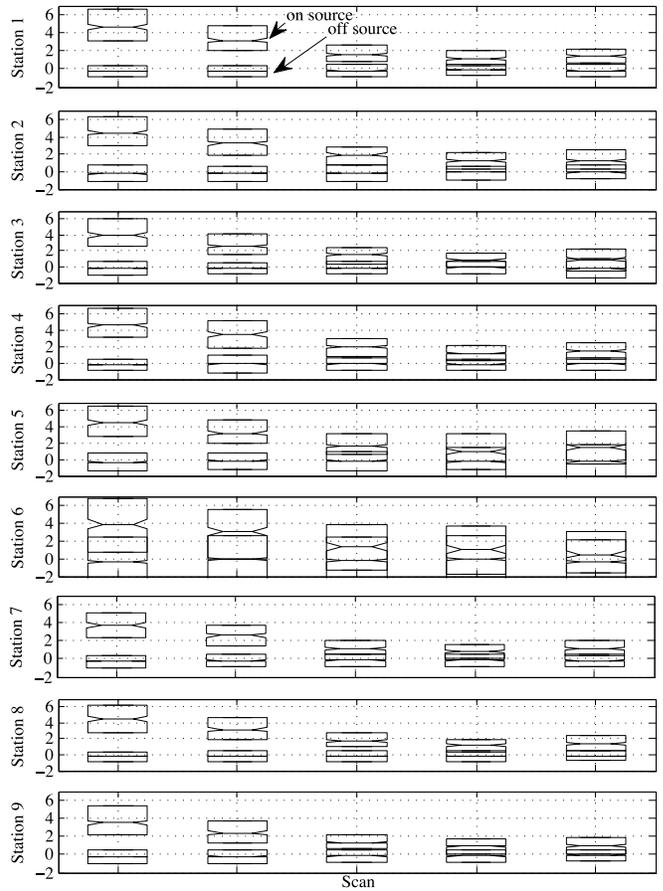, width=\linewidth}}
\caption{Spread of on-source signals (upper distributions) and off-source signals (bottom distributions) for each of nine antennas, over five scans.  The box plots show interquartile ranges, with notches marking the 90\% confidence intervals for the median.  The distributions are better separated during the initial scans.}  
\label{fig:scan_station_quality}
\end{figure}

Figure \ref{fig:scan_station_quality} shows the distribution of received power for pulse and non-pulse segments, grouped by scan and station, and illustrated by top and bottom boxes respectively.  The boxes indicate inter-quartile ranges and medians, with notches marking the 95\% confidence intervals.  Pulse power varies across scans, but these variations seem correlated across stations.  This suggests that the received flux actually changed which favors scintillation as a promising explanation \citep{Rickett1990}.  Reports of scintillation are common in previous studies of pulsar B0329+54.  The cross-scan variability observed here is consistent with the $20$ minute diffractive scintillation cycle observed by \cite{Semenkov2004}. 

\begin{figure}[]
\centerline{\epsfig{file=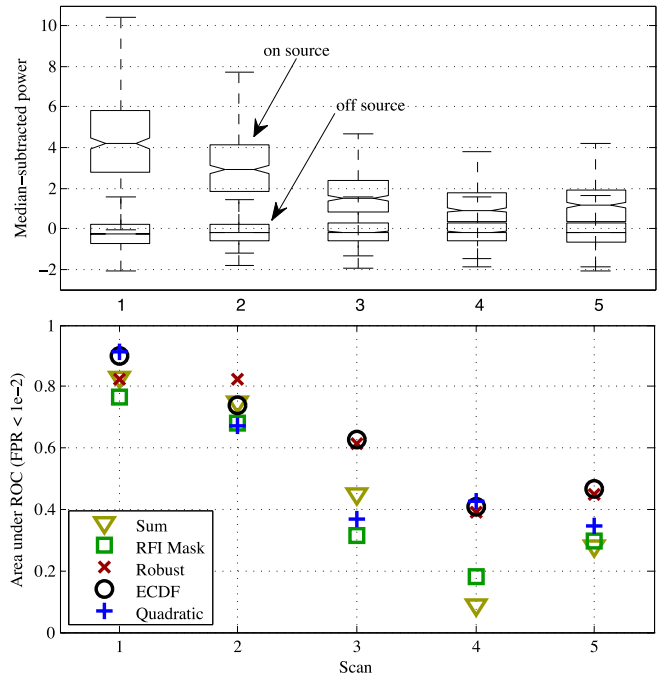, width=\linewidth}}
\caption{Distributions of signal (upper distributions) and off source segments (bottom distributions) for each scan, with boxes corresponding to interquartile ranges and whiskers the extent of the data {\it sans} extreme outliers.  The lower panel shows normalized Areas Under the ROC Curve from Table \ref{tab:roc_pulsar_constrained}.}  
\label{fig:scan_quality}
\end{figure}

Figure \ref{fig:scan_quality} and Table \ref{tab:roc_pulsar_constrained} show the Area Under the ROC Curve scores for each method.  Figure \ref{fig:scan_quality} also compares total on- and off-source power, plotted with box and whisker diagrams in the upper panel.   The differences in signal strength visibly affect performance.  The robust and ECDF discriminants perform best overall due to their ability to discriminate weaker pulses in later scans.  The quadratic discriminant initially performs quite well since the first scan falls directly after its training example when the characteristics of the test data are most similar.  However, this method's performance degrades severely on later scans where the source is weaker relative to RFI.  

\begin{table}[]
\begin{centering}
\begin{tabular}{|l|c|c|c|c|c|c|}
\hline
Method    & Scan 1 & 2      & 3      & 4      & 5   \\
\hline
\hline
Sum & 0.826 & 0.747 & 0.446 & 0.087 & 0.279 \\ 
\hline
Mask & 0.765 & 0.679 & 0.312 & 0.180 & 0.295 \\ 
\hline
Robust &  0.823 & 0.824 & 0.612 & 0.388 & 0.446 \\ 
\hline
ECDF & 0.900 & 0.737 & 0.625 &  0.406 & 0.465 \\ 
\hline
Quadratic & 0.913 & 0.674 & 0.370 & 0.427 & 0.345 \\ 
\hline
\end{tabular}
\end{centering}
\caption{Area Under the ROC Curve (AUC) score of pulsar B0329+54 observations, for realistic false positive rates (less than $0.01$).}
\label{tab:roc_pulsar_constrained}
\end{table}

\begin{table}[]
\begin{centering}
\begin{tabular}{|l|c|c|c|c|c|}
\hline
Method    & Scan 1 & 2      & 3      & 4      & 5      \\
\hline
\hline
Sum       & 0.864 & 0.900 & 0.877 & 0.826 & 0.821 \\
\hline
Mask      & 0.969 & 0.921 & 0.827 & 0.847 & 0.840 \\
\hline
Robust    & 0.814 & 0.920 & 0.866 & 0.899 & 0.881 \\
\hline
ECDF      & 0.807 & 0.936 & 0.850 & 0.879 & 0.895 \\
\hline
Quadratic & 0.846 & 0.834 & 0.904 & 0.892 & 0.845 \\
\hline
\end{tabular}
\end{centering}
\caption{Total Area Under the ROC Curve (AUC) score of pulsar B0329+54 observations, for all false positive rates. }
\label{tab:roc_pulsar}
\end{table}

\begin{figure}[]
\centerline{\epsfig{file=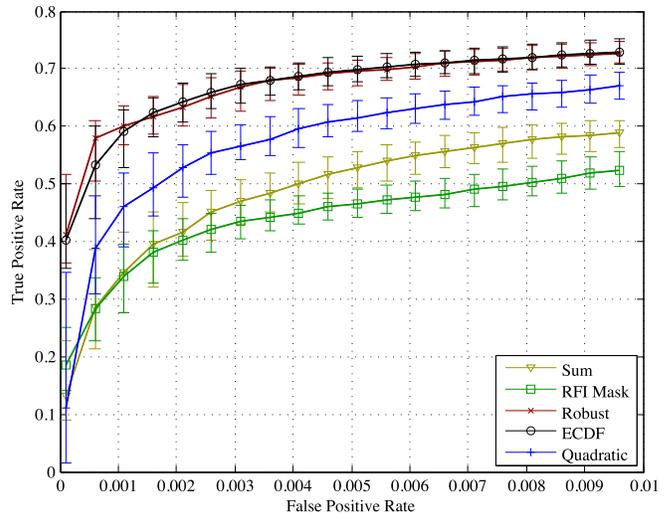, width=\linewidth}}
\caption{Detection performance for the B0329+54 observation, over all scans.  We focus on ROC curve in the relevant region of false positive rates significantly less than 0.01.}  
\label{fig:bm0329}
\end{figure}

We consider the ROC curve in the regime of low tolerances for false positives, and specifically operations-relevant trigger rates that archive no more than 1\% of all candidates (although for completeness, we also report the total AUC scores in Table \ref{tab:roc_pulsar}).  In practice every detection must promote an interval of time around each detection in order to capture the entire dispersed pulse and provide context to characterize RFI.
Figure \ref{fig:bm0329} plots the actual ROC curves of each method up to a $0.01$ promotion rate.  We form confidence intervals for the ROC curve with a bootstrap \citep{Bertail2008}.  Specifically we draw randomized resamplings of the original dataset, recompute classifications and from this the ROC using a kernel-smoothed estimator \cite{Wasserman2006} of true and false positive rates.  Finally, we identify the median ROC curves and 90\% bounding coverage intervals using the bootstrap sample.   

The experiment reveals a highly significant difference between single- and multiple-station approaches.  Multiple-station methods, such the ECDF and robust discriminants, promote more pulses than noise events for similar time budgets.  A realistic budget would permit just a few false positives.  The steep initial slope of multi-station ROC curves implies superior performance in this regime. 

\begin{figure}[]
\centerline{\epsfig{file=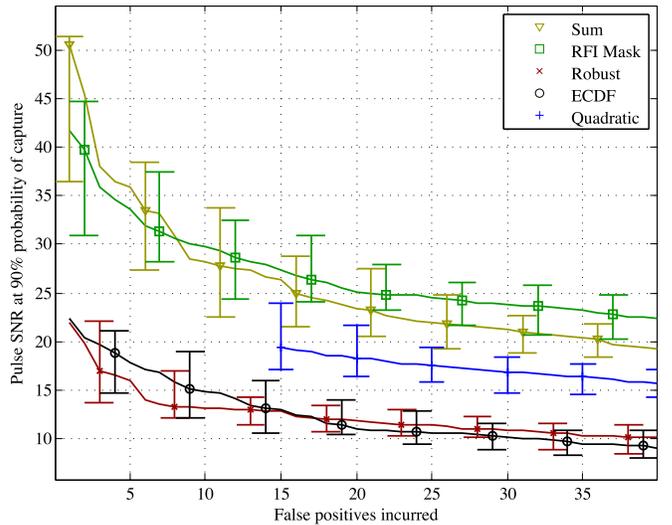, width=\linewidth}}
\caption{Sensitivity of discriminant functions: the SNR of pulses that can be captured with 90\% certainty, for various false positive budgets, over all scans.}  
\label{fig:sensitivity_pulsar}
\end{figure}

Improved ROC performance permits more lenient detection thresholds and improved sensitivity.  Figure \ref{fig:sensitivity_pulsar} shows the SNR of pulses that can be captured by each method for different false positive tolerances.  The SNR associated with an empirical $90\%$ probability of capture is shown, based on the combined dataset from all scans.  We determine 95\% confidence intervals with bootstrap sampling.   Note that there exists a threshold where top performer (the Robust estimator) captures all pulses of SNR $\geq 25$ with $90\%$ probability, without promoting a single non-pulse to archival storage.  The standard summation approach promotes $40$ RFI events before achieving this effective sensitivity. 

\section{Conclusions}

This work demonstrates a case study of real-time incoherent detection of transient signals from multiple stations.   Preliminary tests with the VLBA corroborate our theoretical analysis that uses impulsive noise and homogeneous receivers.   These tests constitute a case study where multiple-station algorithms yields significant performance benefits over a standard summation approach.  When the tolerance for false positives is low, which is the case for most practical installations, multiple-station methods can achieve significant sensitivity improvements without increasing false alarms.   Coupling these techniques with other statistical or multi-band approaches for RFI excision might improve performance further.  For example, alternative RFI mitigation might still be important to excise satellite signals observed simultaneously by multiple stations. 

One unexpected result from the VLBA experiment was that the supervised learning approach (the quadratic decision function) performed worse in practice than our original model predicted.  This discriminant relies on prior examples of both pulses and RFI, so the observed drift in pulse intensity over time invalidates its training assumptions.  Alternatively, stricter regularization could be used to prevent over-fitting of the training data.  Future research may also consider more sophisticated learning algorithms with better generalization properties, such as those that can detect changes in the underlying phenomena (concept drift).  

There are other promising avenues for improvement.  Machine learning techniques can interpret information from features beyond the simple signal measurements used in our tests.  Discriminant functions could incorporate multiple matched filters and dispersion measures.  A natural addition would be to consider the received signal at DM $0$, which is a strong indicator of RFI.  Attempts to expand the feature set should ensure that the resulting discriminant function retain a simple structure to avoid over-fitting to a single training environment and to keep computational requirements tractable for real-time processing.  

\appendix
\section*{Sensitivity of Ensemble Estimators}
\label{app:sensitivity_demonstration}

Here we provide a simple proof sketch that a broad class of multiple-station ensemble detection rules preserves sensitivity.  We consider a discriminant function $q$ that transforms the signal $\phi_i(x,d)$ according to some positive, monotonically increasing, and concave function $r$ at each station independently, and then averages the result across stations.  
\begin{eqnarray}
q(x) &=& \hat{r}(x) = \frac{1}{n}\displaystyle\sum_{a=1}^n r(x_a) \label{eqn:appa_qdef}
\end{eqnarray}
\noindent The ensemble CDF estimator of section \ref{sec:methods} falls into this category insofar as the noise CDF is positive and concave in the region of interest (i.e., larger-than-average values).   We assume a gaussian noise distribution.  The classical detector that averages all $n$ stations yields the noise distribution $\mathcal{N}(\mu,\sigma/\sqrt{n})$.  We aim to show that the ensemble estimator is no less sensitive.  In other words, for some constant on-source signal strength $\tau$:
\begin{eqnarray} 
P \left(q(\mathcal{N}(\mu,\sigma))<q(\tau) \right) &\geq& P(\mathcal{N}(\mu,\sigma/\sqrt{n})<\tau) \label{eqn:appa_goal}\\
P \left(\frac{1}{n}\displaystyle\sum_{a=1}^n r(\mathcal{N}(\mu,\sigma)) < \frac{1}{n}\displaystyle\sum_{a=1}^n r(\tau) \right) &\geq& P(\mathcal{N}(\mu,\sigma/\sqrt{n})<\tau)\nonumber\\ 
P \left(\frac{1}{n}\displaystyle\sum_{a=1}^n r(\mathcal{N}(\mu,\sigma)) < r(\tau) \right)\nonumber &\geq& P(\mathcal{N}(\mu,\sigma/\sqrt{n})<\tau) 
\end{eqnarray}
\noindent For the concave function $r$, Jensen's inequality provides (for some constant $c$):
\begin{eqnarray}
r \left(\frac{1}{n}\displaystyle\sum_{a=1}^n \mathcal{N}(\mu,\sigma)\right) &\geq& \frac{1}{n}\displaystyle\sum_{a=1}^n r(\mathcal{N}(\mu,\sigma))\nonumber \\
P \left(\frac{1}{n}\displaystyle\sum_{a=1}^n r(\mathcal{N}(\mu,\sigma))<c \right) &\geq& P \left(r \left(\frac{1}{n}\displaystyle\sum_{a=1}^n \mathcal{N}(\mu,\sigma)\right)<c\right) \nonumber \\
P(q(\mathcal{N}(\mu,\sigma))<c) &\geq& P \left(r \left(\frac{1}{n}\displaystyle\mathcal{N}(\mu,\sigma) \right)<c \right)\nonumber \\\
P(q(\mathcal{N}(\mu,\sigma))<c) &\geq& P(r(\mathcal{N}(\mu,\sigma/\sqrt{n}))<c) 
\end{eqnarray}
\noindent Therefore, in order to show
\begin{eqnarray}
P(q(\mathcal{N}(\mu,\sigma))<q(\tau)) &\geq& P(\mathcal{N}(\mu,\sigma/\sqrt{n})<\tau) 
\end{eqnarray}
\noindent it is sufficient with transitivity to demonstrate:
\begin{eqnarray}
P(r(\mathcal{N}(\mu,\sigma/\sqrt{n}))<q(\tau)) &\geq& P(\mathcal{N}(\mu,\sigma/\sqrt{n})<\tau)
\end{eqnarray}
\noindent $\tau$ is constant so we can substitute to yield:
\begin{eqnarray}
P(r(\mathcal{N}(\mu,\sigma/\sqrt{n}))<r(\tau)) &\geq& P(\mathcal{N}(\mu,\sigma/\sqrt{n})<\tau)
\end{eqnarray}
\noindent If $r$ is a positive monotonically increasing function, this is a tautology.

\section*{Acknowledgements}
We thank the VLBA administration and operators for their invaluable support, and for access to facilities and hardware that made possible the commensal observations described in this work. The National Radio Astronomy Observatory is a facility of the National Science Foundation operated under cooperative agreement by Associated Universities, Inc.  The V-FASTR project is a trailblazer installation for the Australian Square Kilometre Array Pathfinder's CRAFT fast transients investigation.  ICRAR and Curtin University provided key hardware for the transient detection pipeline, and support for the researchers that participated in the project.  Steven J. Tingay is a Western Australian Premier's Research Fellow.  Dayton Jones and Robert Preston of the Jet Propulsion Laboratory provided key institutional support and guidance for JPL participants.  Peter Hall was a vital liaison to the CRAFT fast transients project.  We also thank J-P Macquart and Sarah Burke-Spolaor for their guidance and insight.  A portion of this research was performed at the Jet Propulsion Laboratory, California Institute of Technology, under a Research and Technology Development Grant.  Copyright 2010.  All Rights Reserved.  US Government Support Acknowledged.
\bibliographystyle{apj}
\bibliography{2010_VFASTR_Detection_Algorithms}
\end{document}